\documentclass[RNAAS]{aastex62}

\begin{document}

\title{A Variable X-ray Source Close to the Magnetar SGR 1935+2154}

%% Note that the corresponding author command and emails has to come
%% before everything else. Also place all the emails in the \email
%% command instead of using multiple \email calls.
\correspondingauthor{Albert Kong}
\email{akong@gapp.nthu.edu.tw}

\author[0000-0002-5105-344X]{A. K. H. Kong}
\affiliation{Institute of Astronomy, National Tsing Hua University, Hsinchu 30013, Taiwan}

\author[0000-0002-0439-7047]{K. L. Li}
\affiliation{Institute of Astronomy, National Tsing Hua University, Hsinchu 30013, Taiwan}

\author{Sangin Kim}
\affiliation{Department of Space Science and Geology, Chungnam National University, Daejeon 34134, Korea}

\author[0000-0003-1753-1660]{C. Y. Hui}
\affiliation{Department of Astronomy and Space Science, Chungnam National University, Daejeon 34134, Korea}

%% Note that RNAAS manuscripts DO NOT have abstracts.
\keywords{stars: individual (SGR 1935+2154) --- stars: magnetars --- stars: neutron --- X-rays: stars}

%% Start the main body of the article. If no sections in the 
%% research note leave the \section call blank to make the title.
\begin{abstract} 
The recent discovery of a millisecond radio burst from the Galactic magnetar \object{SGR 1935+2154} has initiated an intense discussion about the connection between magnetars and fast radio bursts (FRBs). Although some properties of the radio burst from SGR 1935+2154 are not the same as cosmological FRBs, there are theoretical models which propose a connection between magnetars and FRBs (see review by \citealt{2020arXiv200505283M}). In particular, the role of a magnetar wind nebula is included in some models, and therefore it is worthwhile to investigate the X-ray environment of SGR 1935+2154 in more detail. Here, we report on the discovery of an X-ray transient feature near SGR 1935+2154 using archival {\it Chandra} data and discuss its possible origin.
\end{abstract}

\section{Introduction}
The soft gamma-ray repeater (SGR) 1935+2154 is a magnetar with frequent X-ray outbursts since its discovery \citep{2016MNRAS.457.3448I,2020ATel13758....1T}. Although \object{SGR 1935+2154} has not been detected with radio emission and pulsation, a luminous millisecond radio burst \citep{2020arXiv200510324T,2020arXiv200510828B} was found temporally coincident with an X-ray burst \citep{2020arXiv200511071L,2020arXiv200511178R} strongly suggesting that magnetars can produce fast radio bursts (FRBs). At the onset of the recent X-ray activity, a dust scattering X-ray halo was found with {\it Swift} \citep{2020ATel13679....1K,2020arXiv200506335M} and the radio burst was detected within a day. In this note, we report on an investigation of the X-ray environment near SGR 1935+2154.

\section{Data Analysis and Results}
We have examined archival {\it Chandra} images to look for small-scale extended emission. While there are numerous {\it Chandra} observations during the X-ray active states in 2014 and 2016, most of the observations are in timing mode preventing a visual inspection. There are two short ACIS-S observations taken in the imaging mode. We reprocessed ObsID 15874 (PI: Rea) and ObsID 18884 (PI: Kouveliotou) with the most updated calibration files with CIAO. To reduce background contamination, we limited our analysis in the energy range of 0.3--7 keV.

\begin{figure}[h!]
\begin{center}
\includegraphics[scale=0.85,angle=0]{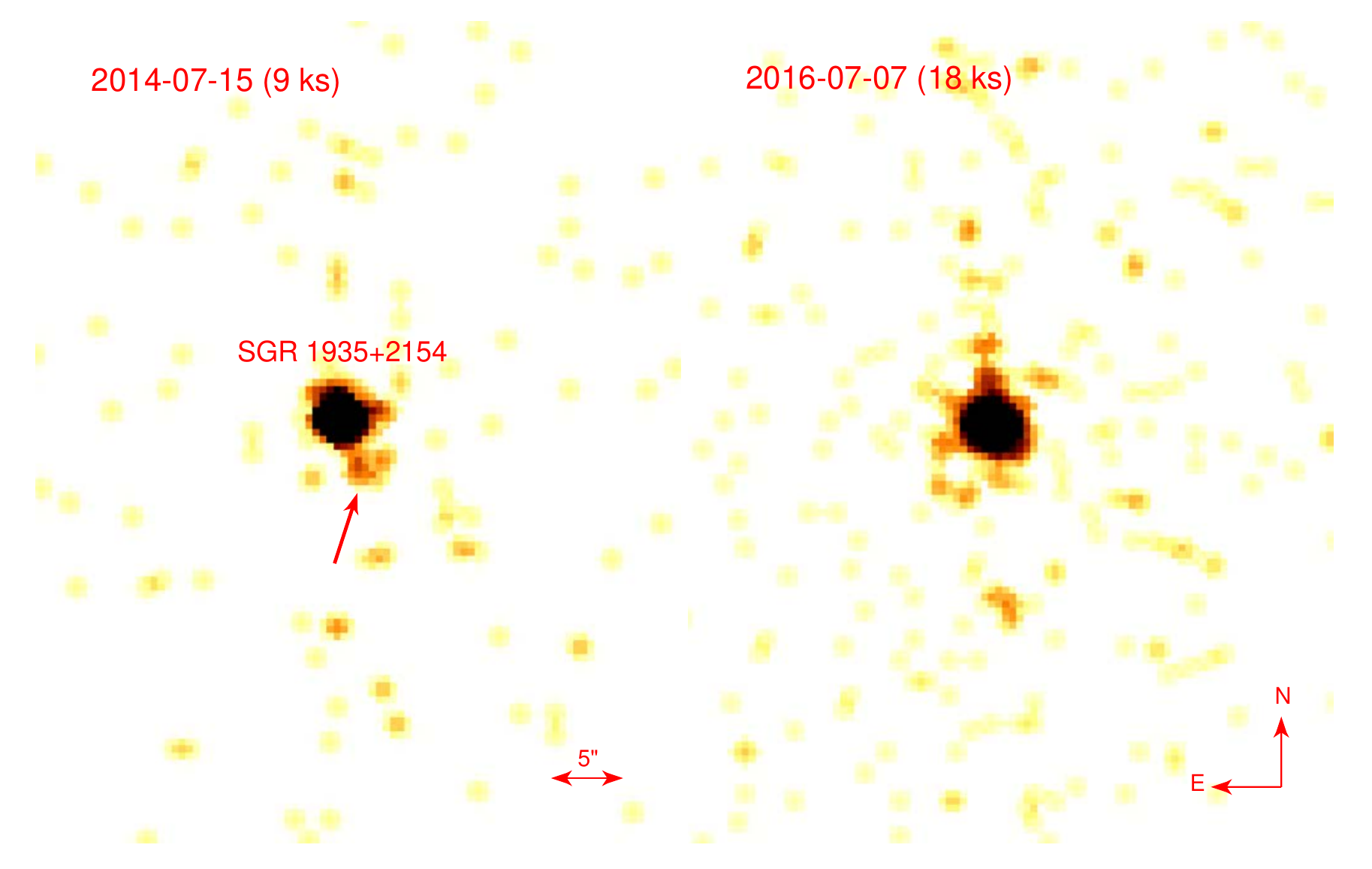}
\caption{{\it Chandra} images of SGR 1935+2154 taken on 2014 July 14 (left) with a feature (marked with an arrow) located at southwest to SGR 1935+2154. The feature disappeared in the image taken on 2016 July 7 (right). Both images were smoothed with a Gaussian with a radius of 2.5 arcsec. \label{fig:1}}
\end{center}
\end{figure}

In Figure \ref{fig:1}, we show the {\it Chandra} images of SGR 1935+2154 taken in 2014 and 2016. In 2014, a faint X-ray source located at about 3 arcsec in the southwest direction from SGR 1935+2154 is seen at $3\sigma$ level. The feature disappeared in the 2016 image which is 2 times deeper. 
We have further examined the radial brightness profiles of SGR 1935+2154 in both observations. Utilizing MARX \citep{2012SPIE.8443E..1AD}, we have simulated the point spread functions (PSFs) of the magnetar and compared that with the observed profiles. In the 2014 observation, the radial profile has a slight deviation from the PSF at $\sim6$~arcsec from the centroid, which is most likely contributed by the aforementioned transient feature. On the other hand, the observed profile in 2016 is consistent with the PSF. Apart from the symmetric radial profile, we have also examined the profile along the N-S and NE-SW orientations. However, we do not find any additional extended feature. 

We next extracted the spectrum of the new source seen in 2014 with a 2 arcsec radius circular region. Since there are only 10 source photons, we employed a Bayesian approach to perform spectral fits using Sherpa \citep{2019JHEAp..21....1L}. With a uniform prior and a Poisson likelihood, we first fitted the spectrum with an absorbed power-law model. In all the following fits, the absorption is fixed at $2.5\times10^{22}$ cm$^{-2}$ based on previous analysis \citep{2017ApJ...847...85Y}. The best-fit photon index $\Gamma$ from the posterior distribution is $6.0\pm1.1$. Since $\Gamma$ is unreasonably large, we next modeled the spectrum with an absorbed blackbody model. The best-fit blackbody temperature is $kT=0.25^{+0.05}_{-0.04}$ keV with an unabsorbed 0.3--7 keV flux of $7.3\times10^{-14}$ erg cm$^{-2}$ s$^{-1}$. 

\section{Discussion}
Here, we discuss the possible origins of the X-ray feature near SGR 1935+2154.

{\it Dust scattering halo:} It is entirely possible that the X-ray feature is part of the halo surrounding SGR 1935+2154. In particular, we expect a soft X-ray spectrum and the transient nature of the feature may be due to the X-ray activities of the magnetar.  However, the last X-ray burst was 10 days before the {\it Chandra} observation \citep{2020ATel13758....1T}. When the feature disappeared in 2016, an X-ray burst was detected 3 days before. Furthermore, instead of a symmetric shape, the feature looks like a point source. We therefore argue that a scattering halo is unlikely.

{\it Supernova remnant:} SGR 1935+2154 is embedded inside the supernova remnant G57.2+0.8 \citep{2016MNRAS.457.3448I}. Given a thermal-like spectrum of the X-ray feature, it is natural to associate it with G57.2+0.8. However, the transient nature argues against this scenario.

{\it Magnetar wind nebula:} It is known that some magnetars have outflows or wind-driven nebulae due to their intense magnetic field. The X-ray spectrum is dominated by synchrotron emission resulting in a power-law shape. The anomalous steep spectrum of the X-ray feature near SGR 1935+2154 is not consistent with our current understanding of magnetar/pulsar wind nebula. The only exception is the pulsar wind nebula around PSR B0656+14 \citep{2016ApJ...817..129B} that has $\Gamma \sim 8$. It has been suggested that such an unusual spectrum might be due to some peculiar gamma-ray properties of the source. On the other hand, X-ray variability of a pulsar wind nebula has been observed, possibly due to shear instabilities or environmental effects \citep{2017ApJ...846..116H}. Therefore, a magnetar wind nebula or outflow may still be possible.

{\it Stellar flare or quiescent X-ray binary in the Galaxy:} The soft spectrum and the corresponding X-ray flux are consistent with a stellar flare or a quiescent neutron star X-ray binary. We examined infrared images provided by the {\it Hubble Space Telescope (HST)} \citep{2018ApJ...854..161L} and did not find any infrared objects down to 26 mag (F140W filter). A non-detection of any infrared objects in the {\it HST} images does not favor this.

{\it Background AGN flare, ultraluminous X-ray source, or tidal disruption event:} The faint transient may be a background event most likely coming from flaring activities of an AGN, an ultraluminosus X-ray source, or a tidal disruption event. The observed X-ray flux implies that the distance to any extragalactic compact objects is very likely less than 200 Mpc. A null detection on the {\it HST} infrared images disfavors this scenario given the distance to the possible host.

In conclusion, the nature of the transient near SGR 1935+2154 remains unclear. Among all scenarios, a magnetar wind nebula or outflow seems more likely although we cannot totally rule out other possibilities. The very steep spectrum and variability might require special conditions in modeling and it is possible that multiple emission components are involved. More {\it Chandra} high-resolution imaging observations are required to shed light on this feature. In particular, if this feature is associated with a magnetar wind nebula of SGR 1935+2154, it would be interesting to check whether it is related to the FRB-like activities. If it is true, magnetar-powered FRB models invoking environmental effects \citep{2020arXiv200505283M} will support the case of SGR 1935+2154.
 
\acknowledgments

This project is supported by the Ministry of Science and Technology of Taiwan via grants 108-2628-M-007-005-RSP and 108-2112-M-007-025-MY3. CYH is supported by the National Research Foundation of Korea through grants 2016R1A5A1013277 and 2019R1F1A1062071. SK is supported by BK21 plus Chungnam National University, National Research Foundation of Korea grants 2016R1A5A1013277 and 2019R1F1A1062071.

%% An example figure call using \includegraphics

\end{document}